# A "HYBRID"-TYPE SYSTEM FOR AN ACHROMATIC FOCUSING OF IONS

B.Yu. Bogdanovich, A.V. Nesterovich,
Moscow Engineering Physics Institute (Technical University)
Kashirskoe sh. 31, Moscow, 115409 Russia

*Abstract*

It is known, that at high energies of ions (more than 10 MeV for light ions) the focusing by magnetic quadrupole fields is most effective, whereas at small energies (less than 1 MeV) - the focusing by electrical quadrupole fields. At intermediate energies of ions (from 1 to 10 MeV) the efficiency of each of these types of a focusing is reduced. The authors offer to utilize for this purpose the "hybrid" type of focusing systems consisting of sequence of electrical and magnetic quadrupole lenses, which are localized in the same space. For making fields of electrical quadrupole lenses it is offered to utilize RF quadrupoles. The RF electrical quadrupoles are created due to the relevant configuration of drift tubes in the vicinity of an accelerating gap. As a result of the general theoretical analysis and calculation of concrete versions of practical realizations it is shown that the given system has advantage as contrasted to by known analogs in the field of medium energies. The velocity band of ions, in which the stable motion is ensured, is essentially changed, and the focusing becomes "achromatic". The calculations have been performed for protons with velocities 0.5-1.0 % of light velocity, at the gradients of electrical field up to 50 kV per square cm, and the gradients of magnetic field up to 1 T/m. Achromaticity of this system allows to lower a radiative activation of accelerator and to reduce damage of electrode surfaces by lost ions.

## 1 INTRODUCTION

In our paper [1], we presented the results of experimental research of energy variable proton linac, which consists of a chain of independently driven one-gap accelerating cavities. A polyaxial resonator (PR) offered as a main accelerating element. In this accelerator, beam focusing has been provided by means of electrostatic quadrupoles with variable potential.

The correct choice of focusing system for such accelerator is very important. At a large variation of accelerator energy, it is necessary to provide focusing for a wide range of particle velocities. It is known, that the focusing by electrical quadrupole (EQ) fields is effective at small energies (less than 1 MeV). At high energies of ions (more than 10 MeV for light ions) the focusing by magnetic quadrupole (MQ) fields is more effective. At intermediate energies of ions (from 1 to 10 MeV) the efficiency of each of these types of a focusing is reduced.

In present paper, the authors offer to utilize the "hybrid" type of focusing systems consisting of sequence of EQ and MQ lenses localized in the same space.

## 2 AN ANALYTICAL ANALYSIS OF A STABILITY DIAGRAM

It is known that in a conventional focusing system consisting of identical MQ's, an operating point on a stability diagram moves from a central part to a boundary at a variation of particle energy. In contrary to this situation, an operating point on a stability diagram of a "hybrid" system moves within a central part of a stability diagram. Let us show it using a simple theoretical model when space regions occupied by MQ lenses and RF EQ's are coincide. This particular case can be realized as drift tubes having RF-electrodes containing permanent magnets. Using notations of Ref. [2], the equations for a transverse motion in this "hybrid" system can be written in the following form

$$v^2 - \gamma^2 = (1-\beta^2)T^2\Omega_q^2, \qquad (1)$$

$$v^2 + \gamma^2 = (1-\beta^2)\sigma T^2\Omega_q^2 \cot\varphi_s + \frac{2Ze}{m_0}\mu_0 T \int_L h(z)dz, \qquad (2)$$

where $\Omega_q^2 = \frac{2\pi ZeE}{\mu_0\beta\lambda}\sin\varphi_s$, $T = \lambda/c$, $\mu_0$ is permeability of free space, $\sigma = \cot(2\pi b/L)$, $\varphi_s$ is a phase of an equilibrium ion (in "cosine" reference system), $\int_L h(z)dz \approx L_{\text{eff}} H/x = L_{\text{eff}} dH/dx = L_{\text{eff}} G$, $\lambda$ is the wavelength in a free space, $c$ is speed of light, $Ze/m_0$ the ration of ion charge to its mass, $\beta$ is a relative ion velocity, $2b/l$ is the gap coefficient, $L_{\text{eff}}$ is an effective length of a magnet quadrupole, $G$ is the gradient of magnetic field of magnet quadrupole, $(v^2-\gamma^2)$ and $(v^2+\gamma^2)$ are the coordinate axes of stability diagram determined according to ref. [2].

To easy a graphical interpretation of our analysis, the equations (1-2) can be rewritten in the following form

$$v^2 - \gamma^2 = D(E/\beta)\sin\varphi_s, \qquad (3)$$

$$v^2 + \gamma^2 = D[B(E/\beta)\cos\varphi_s + C], \qquad (4)$$

where $D = 2\pi Ze\lambda(1-\beta^2)/(m_0 c^2)$, $B = \cot(2\pi b/L)$, $C = c\mu_0 L_{\text{eff}} G/\pi$.

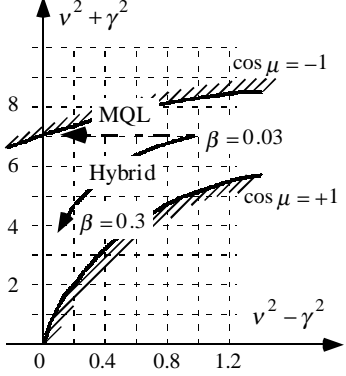

Figure 1: A Typical example for an trajectory of operating point on a stability diagram for a conventional focusing by MQ lenses and by "hybrid" system.

Let us analyze these equations at a constant acceleration rate ($E = $ const), the wavelength of RF fields ($\lambda = $ const) and an identical magnetic quadrupoles ($L_{\text{eff}} = $ const, $G = $ const). It is seen that the term $D$ does not vary at nonrelativistic conditions, i.e. $D \approx $ const.

For a conventional MQ focusing system without RF quadrupoles, the first term in the r.h.s. of eq. (2) is equal to zero. An trajectory of an operating point on a stability diagram with coordinates $\{v^2 - \gamma^2; v^2 + \gamma^2\}$ is parallel to the $(v^2 - \gamma^2)$-axis from a center of the diagram to a boundary.

For a "hybrid" focusing system with RF quadrupoles, the first term in the r.h.s. of eq. (2) reduces at increasing of particle velocity $\beta$. By an optimal choice of parameters, a trajectory of an operating point on a stability diagram can stay within a central part.

Let us consider a numerical example. The equation (4) can be written in the following form

$$v^2 + \gamma^2 = (v^2 - \gamma^2)\cot\varphi_s \cdot \cot(2\pi b/L) + DC. \quad (5)$$

At $\lambda = 1.5\,\text{m}$, $E = 3\,\text{MV/m}$, $\cot\varphi_s = 2$, $\cot(2\pi b/L) = 1.75$, the eq. (5) becomes

$$v^2 + \gamma^2 = 3.5(v^2 - \gamma^2) + DC. \quad (6)$$

For an input of the system with $\beta = 0.03$, let us put $v^2 + \gamma^2 = 7$. Using eq. (6), we obtain $DC = 3.5$. For an output of the system with $\beta = 0.3$, let us put an operating point in a center of a stability diagram, i.e. at $v^2 + \gamma^2 = 4$. It can be shown the maximum value of magnetic field at a magnet surface must be equal to 1.3 T for protons. It can be realized for permanent magnets made from samarium-cobalt alloy. Trajectories of operating points on a stability diagram are shown in Fig.1.

## 3 A NUMERICAL MATRIX ANALYSIS OF A REAL SYSTEM

Let us analyze a real system when MQ lenses and RF EQ lenses occupy different space regions. It can be performed using transfer matrices [2-4]. The notation adopted in Ref.[3,4] is used for a description of the transverse motion in periodic quadrupole fields. The linear motion of particles is described by the following equation for dimensionless displacements, $x$ from the beam axis, $z$

$$\begin{cases} d^2x/dz^2 + P(z)x = 0, \\ d^2y/dz^2 + \overline{P}(z)y = 0, \end{cases} \quad (7)$$

where $P(z) = -A(\varphi)g(z) + \Lambda^2 h(z)$ and $\overline{P}(z) = -A(\varphi)g(z) - \Lambda^2 h(z)$ are real functions describing an structure of an accelerating-focusing channel. The dimensionless coefficients $A(\varphi)$ and $\Lambda^2$ describe defocusing and focusing effects, respectively. They are

$$A(\varphi) = \frac{\pi e E_m (1-\beta_s^2)^{3/2} L^2}{m_0 c^2 \lambda \beta^3} \sin\varphi, \quad (8)$$

$$\Lambda^2 = \frac{eH'\sqrt{1-\beta_s^2}L^2}{m_0 c^2 \beta_s} = \frac{eE'\sqrt{1-\beta_s^2}L^2}{m_0 c^2 \beta_s^2}, \quad (9)$$

where $E'$ and $H'$ are gradients of electrical and magnetic quadrupole lenses, respectively, $E_m$ is an amplitude of an accelerating field, $g(z)$ and $h(z)$ are real functions describing a distribution of defocusing and focusing forces.

For matrix analysis of the transverse motion, it is necessary to know matrices describing elementary sections of an accelerating-focusing channel. In our case, the matrices for a drift space $a_{\text{drift}}$, accelerating gap $a_{\text{ag}}$, defocusing and focusing quadrupole $a_{\text{dq}}$ and $a_{\text{fq}}$ are used. Let us assume that defocusing forces are concentrated in a gap center [3]. Then the matrices are expressed in the following form

$$a_{\text{drift}} = \begin{bmatrix} 1 & \varepsilon_d/N \\ 0 & 1 \end{bmatrix}, \quad a_{\text{fq}} = \begin{bmatrix} \cos q & \sin q/\Lambda \\ -\Lambda \sin q & \cos q \end{bmatrix},$$

$$a_{\text{ag}} = \begin{bmatrix} 1 & D \\ A/N & 1 \end{bmatrix}, \quad a_{\text{dq}} = \begin{bmatrix} \cosh q & \sinh q/\Lambda \\ \Lambda \sinh q & \cosh q \end{bmatrix}, \quad (10)$$

where $q = \Lambda \varepsilon_q/N$, and $\varepsilon_d$, $\varepsilon_q$ are the normalized length of sections, $N$ is the number of gaps within a channel period.

Then matrices describing elementary sections are multiplied and a resulting transfer matrix $[a_{ij}]$ is analyzed according to the known condition of motion stability [3]: $-1 < (a_{11} + a_{22})/2 < 1$.

Stability of a transverse motion has been analyzed for an accelerating-focusing channel shown in Fig.2 at

$H' = 1$ T/cm, $E' = 50$ kV/cm, and $\beta_p = 0.05$. We have considered two systems. The first one is a conventional MQ focusing system without RF quadrupoles and the second one is the proposed "hybrid"-type systems. RF EQ field is provided by turning of elliptical apertures of neighboring drift tubes by $90^o$. Parameters of these systems are presented in Table 1 and 2.

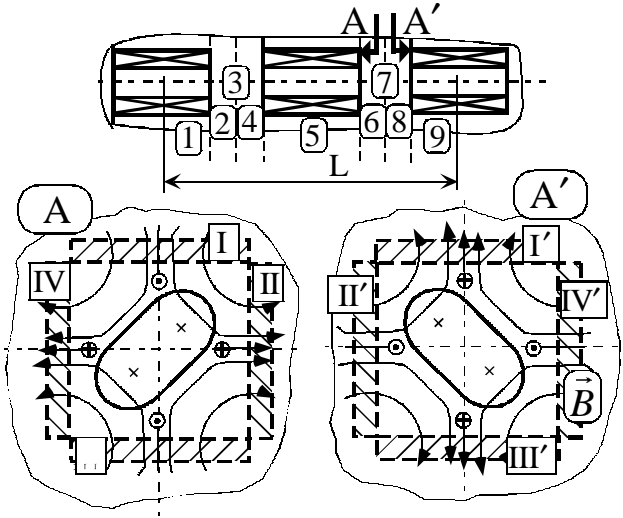

Figure 2: A general view of accelerating-focusing channel and its partitioning into elementary sections for a matrix analysis (upper drawing), and an example of a technical realization of "hybrid"-type focusing with numbering of permanent magnets (lower drawings).

A "hybrid"-type focusing provides an additional degree of freedom for an optimization of focusing parameters. The gradients of electrical and magnetic lens can be varied from zero to a maximum value.

At identical lengths of accelerating gaps, view of a stability diagram will change due to a reduction of a focusing parameter for RF electrical lenses, because $\Lambda^2 \propto E'/\beta$. For example, at $\beta = 0.005$, the stability diagram is near the same as a diagram when magnetic lenses are absent. At increasing $\beta$, a stability diagram becomes close to a diagram for magnetic quadrupole system without RF quadrupoles. Figure 3 shows stability diagrams of a conventional magnetic-quadrupole focusing system and a stability diagram of a "hybrid"-type system at different $\beta$. The focusing parameter of magnetic quadrupoles is used as an ordinate.

An application of the "hybrid"-type system extends a velocity of focused ions without a change of focusing system parameters. It provides a possibility of optimization of an accelerator in terms of a radial motion.

## ACKNOWLEDGEMENTS
The authors express thanks to Mr. V.Kuevda for help with numerical calculations.

Table 1: Parameters of a conventional magnetic-quadrupole focusing system without RF quadrupoles

| Section No. | Type of section | The section length, cm |
|---|---|---|
| 1 | 1/2 of focusing lens | 3 |
| 2 | Drift space | 1 |
| 3 | Thin-gap | 0 |
| 4 | Drift space | 1 |
| 5 | Defocusing lens | 6 |
| 6 | Drift space | 1 |
| 7 | Thin-gap | 0 |
| 8 | Drift space | 1 |
| 9 | 1/2 of focusing lens | 3 |

Table 2: Parameters of a "hybrid"-type system

| Section No. | Type of section | Length, cm |
|---|---|---|
| 1 | 1/2 of magnetic foc. lens | 3 |
| 2 | Electrical defocusing lens | 1 |
| 3 | Thin-gap | 0 |
| 4 | Electrical focusing lens | 1 |
| 5 | magnetic defocusing lens | 6 |
| 6 | Electrical focusing lens | 1 |
| 7 | Thin-gap | 0 |
| 8 | Electrical defocusing lens | 1 |
| 9 | 1/2 of magnetic foc. lens | 3 |

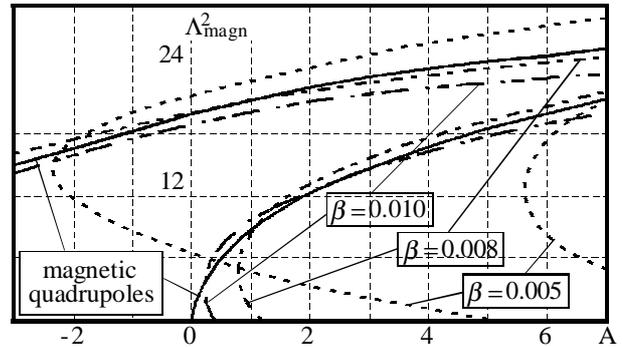

Figure 3: The stability diagram for a conventional magnetic quadrupoles and "hybrid"-type system.